\documentclass[twocolumn,prd,superscriptaddress,showpacs,floatfix%
,preprintnumbers,nofootinbib]{revtex4}

\usepackage{graphicx}

\begin{document}

\title{Solar neutrino limit on axions and keV-mass bosons}

\author{Paolo Gondolo}
\affiliation{Department of Physics, University of Utah,
115 S 1400 E \#201, Salt Lake City, UT 84102, USA}

\author{Georg~G.~Raffelt}
\affiliation{Max-Planck-Institut f\"ur Physik
(Werner-Heisenberg-Institut), F\"ohringer Ring 6, 80805 M\"unchen,
Germany}

\date{18 July 2008, revised 21 April 2009}

\preprint{MPP-2008-69}
\begin{abstract}
The all-flavor solar neutrino flux measured by the Sudbury Neutrino
Observatory (SNO) constrains nonstandard energy losses to less than
about 10\% of the Sun's photon luminosity, superseding a
helioseismological argument and providing new limits on the
interaction strength of low-mass particles. For the axion-photon
coupling strength we find $g_{a\gamma}<7\times10^{-10}~{\rm
GeV}^{-1}$. We also derive explicit limits on the Yukawa coupling to
electrons of pseudoscalar, scalar and vector bosons with keV-scale
masses.
\end{abstract}


\maketitle

\section{Introduction}                        \label{sec:introduction}

The interaction strength of new low-mass particles with photons,
electrons or nucleons is severely constrained by the well-known
requirement that stars not lose energy in excess of observational
constraints~\cite{Raffelt:1990yz, Raffelt:1996wa, Raffelt:2006cw}.
Depending on the assumed particle mass and interaction structure,
the most restrictive limits typically derive from population
statistics in globular clusters, the white dwarf luminosity
function, and the duration of the SN~1987A neutrino signal. Such
constraints usually imply that the solar particle emission is small
compared to its photon luminosity, but it can still serve as a
powerful source, e.g.\ for the ongoing solar axion
searches~\cite{Moriyama:1998kd, Inoue:2008zp, Andriamonje:2007ew,
Arik:2008mq}.

Constraints based on the properties of the Sun remain of interest,
even if they are less restrictive than other astrophysical
arguments, because they are more direct and thus perhaps more
comparable to laboratory experiments. Moreover, sometimes the Sun is
enough to test a given hypothesis. In this case it is nice to have a
simple argument at hand that does not require more complicated
astrophysical reasoning.

Previously the most sensitive diagnostic for the solar interior was
the helioseismological sound-speed profile, providing restrictive
energy-loss limits~\cite{Schlattl:1998fz}. Of course, the underlying
chain of arguments is not simple and worse, a new determination of
the solar element abundances~\cite{Asplund:2004eu} has created an
unresolved tension between solar modeling and
helioseismology~\cite{Bahcall:2004yr, PenaGaray:2008qe}. While the
conservative limit of Ref.~\cite{Schlattl:1998fz} likely remains
unchanged, it is nice that the measured solar neutrino flux provides
a somewhat more restrictive limit based on a simpler argument.

The recently completed SNO measurements of the all-flavor solar
neutrino flux~\cite{Ahmad:2002jz, Aharmim:2005gt, Aharmim:2008kc}
probe directly the physical conditions of the particle-emitting
solar core. The steep temperature dependence of the $^8$B neutrino
production rate provides a sensitive test of the Sun's interior that
would be hotter if a lot of ``invisible energy'' were produced. The
main purpose of our note is to update solar energy-loss constraints
using the SNO results.

The inner solar temperature is around 1~keV, so these limits always
apply to sub-keV mass particles, notably axions. Recently the
hypothesis of keV-scale bosons as possible dark matter candidates has
received some attention~\cite{Pospelov:2008jk}. Moreover, some time
ago it was proposed~\cite{Bernabei:2005ca} that keV-mass pseudoscalars
explain the annual modulation observed in the DAMA/Libra
experiment~\cite{Bernabei:2008yi}. Unfortunately, this intriguing
intepretation was based on the incorrect axio-electric absorption rate
of Refs.~\cite{Bernabei:2005ca, Dimopoulos:1986mi}, the correct rate
being much larger but nearly independent of
velocity~\cite{Pospelov:2008jk}. While an earlier version of our
manuscript was largely motivated by this now-dismissed interpretation,
an evaluation of the solar limit for keV-mass bosons may still prove
useful in future.

We present the new solar energy-loss constraint in the context of
axions and apply it explicitly to the axion-photon interaction
strength. This result is of interest for solar axion searches. We
also treat explicitly keV-mass bosons $\chi$ that couple to
electrons. To this end we re-derive the $\gamma+e^-\to e^-+\chi$
Compton cross section, correcting several errors in the literature.

\section{Axion-photon interaction}            \label{sec:axion-photon}

Axions are of particular interest because the Sun is used as a
source for ongoing helioscope searches~\cite{Moriyama:1998kd,
Inoue:2008zp, Andriamonje:2007ew, Arik:2008mq}. Axions are produced
by the Primakoff process $\gamma+Ze\to Ze+a$, that is mediated by a
virtual photon due to the axion's two photon interaction ${\cal
L}_{a\gamma}=-\frac{1}{4}g_{a\gamma} F_{\mu\nu}\tilde F^{\mu\nu}a=
g_{a\gamma}{\bf E}\cdot{\bf B}a$. In the laboratory, solar axions
convert back into $X$-rays while traveling along a dipole magnet
oriented toward the Sun. For $m_a\alt 0.2$~eV,
CAST~\cite{Andriamonje:2007ew} provides the most restrictive limit
of $g_{10}<0.88$ at 95\% CL, where $g_{10}=g_{a\gamma}/10^{-10}~{\rm
GeV}^{-1}$. The stellar energy-loss limit from globular-cluster
stars is comparable, but without a detailed budget of systematic
uncertainties.

The axion luminosity $L_a=g_{10}^2\,1.85\times10^{-3}\,L_\odot$
\cite{Andriamonje:2007ew} represents a negligible perturbation of
the Sun if $g_{10}$ is below the CAST limit. However, for larger
couplings the energy loss modifies the solar structure. To maintain
the observed amount of energy emitted at the surface, more energy
than usual needs to be produced by nuclear burning. The latter is
self--regulating, so the energy-producing regions must heat up. The
extra losses would have operated for the entire lifetime of the Sun
so that one must evolve a zero-age model to its present age of
\hbox{$4.6\times10^9$ years}, at which point it must match the
present-day radius and surface luminosity. One adjusts the unknown
pre-solar helium abundance to achieve this~fit.

Schlattl et al.\ (1998) have produced a series of such
self-consistent present-day solar models for different levels of
axion emission based on the Primakoff effect~\cite{Schlattl:1998fz}.
They provide the required pre-solar helium abundance and show the
present-day central helium abundance, density and temperature as
well as the neutrino fluxes. In 1998 the question of neutrino flavor
oscillations was not yet settled. Therefore, Schlattl et al.\ used
helioseismology to provide a conservative constraint
$L_a<0.20\,L_\odot$, corresponding to $g_{10}\alt10$.

The all-flavor solar neutrino flux from the $^8$B reaction measured
by the SNO experiment~\cite{Ahmad:2002jz, Aharmim:2005gt,
Aharmim:2008kc} is a more direct probe. For $L_a\alt 0.5\,L_\odot$
the self-consistent solar models of Schlattl et
al.~\cite{Schlattl:1998fz} provide with excellent accuracy
\begin{eqnarray}\label{eq:boronflux}
\Phi_{\rm B8}^a&=&\Phi_{\rm B8}^0\,
\left(\frac{L_\odot+L_a}{L_\odot}\right)^{4.6}\,,\\*
T_{\rm c}^a&=&T_{\rm c}^0\,
\left(\frac{L_\odot+L_a}{L_\odot}\right)^{0.22}\,,
\end{eqnarray}
where $\Phi_{\rm B8}^a$ is the $^8$B solar neutrino flux for a solar
model with axion losses $L_a$ whereas $\Phi_{\rm B8}^0$ is for the
standard case, and similar for the central temperature~$T_{\rm c}$.

These power laws follow from a simple scaling argument because we
are in a regime where the axion flux is a small perturbation. The
second equation shows that energy generation by hydrogen burning for
solar conditions scales approximately with $T^{4.5}$ and the $^8$B
flux varies roughly as $T^{18}$. The main advantage of
Eq.~(\ref{eq:boronflux}) is that it uses the constraint of a
self-consistent present-day solar model and that one has a direct
connection between the Sun-averaged neutrino and axion fluxes.

The all-flavor solar neutrino flux from the $^8$B reaction was
measured by SNO. The pure D$_2$O phase provided a flux of
$5.09^{+0.44}_{-0.43}$(stat)$^{+0.46}_{-0.43}$(sys) in units of
$10^6~{\rm cm}^{-2}~{\rm s}^{-1}$ \cite{Ahmad:2002jz}. The salt
phase provided $4.94^{+0.21}_{-0.21}$(stat)$^{+0.38}_{-0.34}$(sys)
\cite{Aharmim:2005gt}. Very recently, the $^3$He phase gave
$5.54^{+0.33}_{-0.31}$(stat)$^{+0.36}_{-0.34}$(sys)
\cite{Aharmim:2008kc}. The old solar models predicted 5.94 in the
same units, whereas the new opacities lead to~4.72, each with a
nominal $1\sigma$ error of 11\% \cite{PenaGaray:2008qe}. The main
non-abundance contributions to this uncertainty are opacity (6.8\%),
diffusion (4.2\%) and the $S_{17}$ factor for the $p+{}^7$Be
reaction (3.8\%).

The measurements and predictions agree well within the stated
errors, although the dominant uncertainty of the calculated fluxes
evidently is from the assumed element abundances. It appears
reasonably conservative to assume the true neutrino flux does not
exceed the prediction by more than 50\%\ so that
\begin{equation}\label{eq:losslimit}
L_{a}<0.1\,L_\odot\,.
\end{equation}
This nominal limit implies
\begin{equation}\label{eq:solarlimit}
g_{a\gamma}<7\times10^{-10}~{\rm GeV}^{-1}\,,
\end{equation}
somewhat more restrictive than the helioseismological limit. The
Tokyo helioscope search provides a limit very similar to this
result~\cite{Moriyama:1998kd, Inoue:2008zp}, whereas the CAST search
is significantly more sensitive~\cite{Andriamonje:2007ew,
Arik:2008mq} and therefore self-consistent: An axion flux on the
level of the CAST limit would not cause any other observable
modification of the Sun or of the solar neutrino flux.

The sensitivity of the helioscope technique quickly diminishes for
$m_a\agt1$~eV. An alternative is Bragg conversion in the strong
electric field within a crystal lattice. This approach extends to
keV-scale masses because the spatial $E$-field variation in the
crystal provides the required momentum difference. Constraints on
$g_{a\gamma}$ from such experiments~\cite{Cebrian:1998mu,
Avignone:1997th, Morales:2001we, Bernabei:2001ny,
Collaboration:2009ht} are however less restrictive than the solar
limit of Eq.~(\ref{eq:solarlimit}). The most recent constraint from
the CDMS experiment is $g_{a\gamma}<24\times10^{-10}~{\rm GeV}^{-1}$
at 95\%~CL for $m_a<0.1$~keV \cite{Collaboration:2009ht}.

\section{Boson-electron coupling}
\label{sec:axion-electron}

The exact energy-loss mechanism is irrelevant for the limit of
Eq.~(\ref{eq:losslimit}) even though the spatial distribution of
particle emission somewhat depends on the temperature and density
variation of the relevant emission process. So we may consider other
reactions besides the axion Primakoff process.

A case in point motivated by the hypothesis of keV scale dark
matter~\cite{Pospelov:2008jk} are bosons $\chi$ that interact with
electrons through a Yukawa coupling $g_{\chi e e}$. Such particles
are emitted from stars by bremsstrahlung $e+Ze\to Ze+e+\chi$ and the
Compton process $\gamma+e\to e+\chi$. For pseudoscalars,
bremsstrahlung contributes about 75\% of the total emission in the
Sun, Compton about 25\% \cite{Raffelt:1985nk}. However, the energy
spectrum for bremsstrahlung is much softer than for Compton. For keV
mass particles threshold effects are important, so it is enough to
use the Compton process alone.

We have calculated the Compton cross sections for the pseudoscalar
(PS), scalar (S), and vector (V) cases for bosons with a nonzero
mass $m_\chi$. The interaction is
\begin{equation}
{\cal L}_{\chi e e}=g_{\chi e e}\times
\cases{i\chi\overline{e}\gamma_5e&PS,\cr
\chi\overline{e}e&S,\cr
\chi_\mu\overline{e}\gamma^\mu e&V.\cr}
\end{equation}
General expressions for the total Compton cross section are given in
the Appendix, superseding for PS an erroneous result in the
literature~\cite{Mikaelian:1978jg}.\footnote{In
  Ref.~\cite{Mikaelian:1978jg} the factors of 2 in the argument of the
  logarithm in Eq.~(\ref{eq:Compton}) are missing. In an earlier
  version of our paper we had used this incorrect expression
  and found a spurious cross-section
  increase with $m_\chi$. Our limits would have excluded the full
  range of PS parameters explaining the DAMA annual
  modulation~\cite{Bernabei:2005ca}, an interpretation that itself was
  based on a spurious cross section~\cite{Pospelov:2008jk}.}  For the
application in the Sun we take the limit of nonrelativistic electrons
with mass $m_e\gg\omega$ (photon energy) and use the velocity of the
outgoing $\chi$ boson $\beta=\sqrt{1-(m_\chi/\omega)^2}$ to express
the cross sections. For~PS we find
\begin{equation}
\sigma_{\rm PS}=\frac{g_{\chi e e}^2\alpha}{3m_e^2}
\,\frac{\omega^2}{m_e^2} \, \frac{\beta (3+\beta^4)}{4}
\end{equation}
in agreement with Ref.~\cite{Pospelov:2008jk}. This is a
superposition of a final-state $s$ and $d$ wave. For the other cases
we find
\begin{equation}
 \sigma=\frac{g_{\chi e e}^2\alpha}{3m_e^2}\times
 \cases{\beta^3&S,\cr
 \beta(3-\beta^2)&V.\cr}
\end{equation}
For S this is a final-state $p$ wave, for V a superposition of $s$
and $p$. For a massless $\chi$ boson we have $\beta=1$ and the V
cross section is twice that of S, representing 2 interacting spin
degrees of freedom. For the other extreme $\beta\to0$ our result
reflects 3 interacting degrees of freedom relative to~S. In
Ref.~\cite{Pospelov:2008jk} the V cross section was stated without
the velocity factors.

\begin{figure}[t]
\includegraphics[width=0.9\columnwidth]{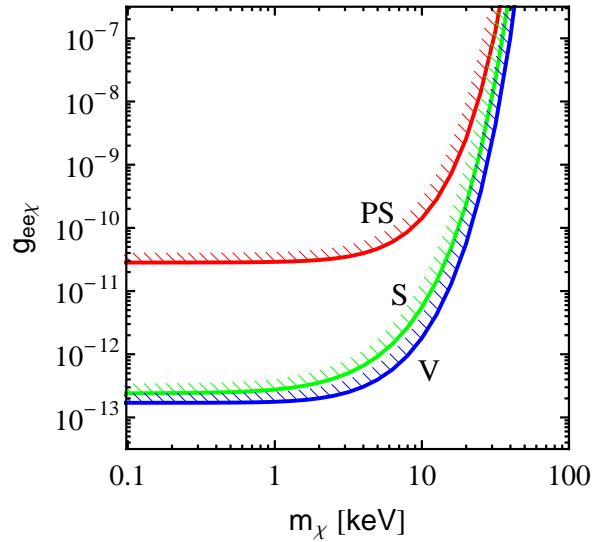}
\caption{Mass dependence of the solar $g_{ee\chi}$ limit. The PS
curve supersedes the one in our original
preprint that was reproduced in
Ref.~\cite{Aalseth:2008rx}.\label{fig:exclusion}}
\end{figure}

We integrate the emission rate over a standard solar
model~\cite{Bahcall:2004pz} and find explicitly for $m_\chi=0$
\begin{equation}
 L_{\chi}^{\rm Compton}=g_{ee\chi}^2\times
 \cases{1.25\times10^{20}\,L_\odot&PS,\cr
 1.72\times10^{24}\,L_\odot&S,\cr
 3.44\times10^{24}\,L_\odot&V.\cr}
\end{equation}
With Eq.~(\ref{eq:losslimit}) this implies the constraints
\begin{equation}
 g_{ee\chi}<\cases{2.8\times10^{-11}&PS,\cr
 2.4\times10^{-13}&S,\cr
 1.7\times10^{-13}&V.\cr}
\end{equation}
We show the $m_\chi$ dependence of these limits in
Fig.~\ref{fig:exclusion}.

\section{Summary}
\label{sec:summary}

The SNO measurements of the all-flavor solar neutrino flux produced
by the very temperature-dependent ${}^8$B reaction severely
constrain anomalous solar energy losses. We have re-considered
self-consistent solar models produced by Schlattl
et~al.~\cite{Schlattl:1998fz} who included axion losses by the
Primakoff effect. We have observed that the predicted solar neutrino
flux is nicely reproduced by a simple and intuitive power law as a
function of the assumed anomalous solar energy loss. In this way the
measured neutrino flux and the assumed energy loss are directly
related in a simple form. The excellent agreement between the
measured and predicted solar neutrino flux provides a restrictive
limit on any new energy loss channel of the Sun. While constraints
from other astrophysical arguments are usually more restrictive, the
solar neutrino limit on new energy losses is complementary in that
it is based on a direct diagnostic of the solar interior.

In particular, we have derived a new solar limit on the axion-photon
interaction strength $g_{a\gamma}$,  superseding an often-cited
helioseismological result. Only the CAST experiment is sensitive
enough to detect solar axions obeying our new constraint
Eq.~(\ref{eq:solarlimit}).

For bosons coupling to electrons, our limit extends to masses of
almost 10~keV even though the solar inner temperature is around
1~keV. This would have been of interest to constrain the DAMA annual
modulation in terms of keV-scale pseudoscalar dark matter particles.
However, based on the corrected axio-electric absorption rate of
Ref.~\cite{Pospelov:2008jk} this interpretation is no longer viable.
Instead, recent direct constraints on keV-scale pseudoscalar dark
matter by CoGeNT~\cite{Aalseth:2008rx} and
CDMS~\cite{Collaboration:2009ht} are more restrictive than the solar
limit.

\begin{acknowledgments}
We acknowledge an illuminating correspondence with M.~Pospelov on
the axio-electric effect and with J.~Collar on the CoGeNT results.
P.G.~acknowledges support from NSF grant PHY-0456825.
G.R.~acknowledges partial support by the Deutsche
Forschungsgemeinschaft under Grant TR-27 ``Neutrinos and Beyond'',
by The Cluster of Excellence ``Origin and Structure of the
Universe.''
\end{acknowledgments}


\appendix*

\section{Cross sections \label{sec:appendix}}

For future reference, we list here the complete expressions of the
Compton cross sections for the production of massive pseudoscalar,
scalar, and vector bosons. We use the notation of
Ref.~\cite{Mikaelian:1978jg}, namely
$p_0=(s-m_e^2+m_\chi^2)/2\sqrt{s}$, $p=(p_0^2-m_\chi^2)^{1/2}$,
$k_0=(s+m_e^2)/2\sqrt{s}$, and $k=\sqrt{s}-k_0$. We find
\begin{equation}\label{eq:Compton}
\sigma=\frac{\alpha g_{\chi ee}^2}{8s} \frac{p}{k} \left[ A(s) + B(s) \frac{\sqrt{s}}{p} \log \frac{2p_0k_0+2pk-m_\chi^2}{2p_0k_0-2pk-m_\chi^2} \right],
\end{equation}
where
\begin{equation}
A(s) = \cases{
-3+\frac{m_e^2-m_\chi^2}{s}+\frac{8m_\chi^2s}{(s-m_e^2)^2},&PS,\cr
-3+\frac{m_e^2-m_\chi^2}{s}+\frac{8(m_\chi^2-4m_e^2)s}{(s-m_e^2)^2}, &S,\cr
2+\frac{2 (m_e^2-m_\chi^2)}{s}+\frac{16 (m_\chi^2+2m_e^2)s}{(s-m_e^2)^2}, &V,}
\end{equation}
and
\begin{equation}
B(s) = \cases{
1 - \frac{2m_\chi^2}{s-m_e^2}+ \frac{2m_\chi^2(m_\chi^2-2m_e^2)}{(s-m_e^2)^2}, & PS,\cr
1 + \frac{2(m_\chi^2-4m_e^2)}{s-m_e^2}+ \frac{2(m_\chi^4-6m_\chi^2m_e^2+8m_e^4)}{(s-m_e^2)^2}, &S,\cr
2-\frac{4 (m_\chi^2+2 m_e^2)}{s-m_e^2}-\frac{4 (4 m_e^4-m_\chi^4)}{\left(s-m_e^2\right)^2}, &V.}
\end{equation}
For $m_\chi=0$ and $g_{\chi ee}^2=4\pi\alpha$, the V cross section
reduces to the usual Compton cross section for $\gamma + e \to e +
\gamma$.


\end{document}